\documentclass[letterpaper, 10 pt, conference]{ieeeconf} 
\IEEEoverridecommandlockouts                              

\usepackage{cite}
\usepackage{amsmath,amssymb,amsfonts}
\usepackage{algorithmic}
\usepackage{graphicx}
\usepackage{textcomp}
\usepackage{xcolor}
\usepackage{siunitx}
\usepackage{amsmath}
\usepackage{url}
\def\BibTeX{{\rm B\kern-.05em{\sc i\kern-.025em b}\kern-.08em
    T\kern-.1667em\lower.7ex\hbox{E}\kern-.125emX}}

\title{\LARGE \bf
\textit{Profi-Load}: An FPGA-Based Solution for Generating Network Load in  Profinet Communication
}

\author{Ahmad Khaliq, Sangeet Saha, Bina Bhatt, Dongbing Gu and Klaus McDonald-Maier
\thanks{This work is funded by INTERREG V 2 SEAS PROJECT INCASE 2S01-049.}
	\thanks{All authors are with the Embedded and Intelligent Systems Laboratory in Computer Science and Electronic Engineering department,
		University of Essex, Colchester, United Kingdom.
		{\tt\small \{ahmad.khaliq,sangeet.saha,bb18131,dgu,kdm\} @essex.ac.uk}}%
		}
				
\begin{document}
\maketitle

\begin{abstract}
Industrial automation has received a considerable attention in the last few years with the rise of Internet of Things (IoT). Specifically, industrial communication network technology such as Profinet has proved to be a major game changer for such automation. However, industrial automation devices often have to exhibit robustness to dynamically changing network conditions and thus, demand a rigorous testing environment to avoid any safety-critical failures. Hence, in this paper, we have proposed an FPGA-based novel framework called ``\textit{Profi-Load}'' to generate Profinet traffic with specific intensities for a specified duration of time. The proposed \textit{Profi-Load} intends to facilitate the performance testing of the industrial automated devices under various network conditions. By using the advantage of inherent hardware parallelism and re-configurable features of FPGA, \textit{Profi-Load} is able to generate Profinet traffic efficiently. Moreover, it can be reconfigured on the fly as per the specific requirements. We have developed our proposed \textit{Profi-Load} framework by employing the Xilinx-based ``NetJury" device which belongs to Zynq-7000 FPGA family. A series of experiments have been conducted to evaluate the effectiveness of \textit{Profi-Load} and it has been observed that \textit{Profi-Load} is able to generate precise load at a constant rate for stringent timing requirements. Furthermore, a suitable Human Machine Interface (HMI) has also been developed for quick access to our framework. The HMI at the client side can directly communicate with the NetJury device and parameters such as, required load amount, number of packet(s) to be sent or desired time duration can be selected using the HMI.  
\end{abstract}


\section{Introduction}
A significant increase in the demand for Ethernet based infrastructure has been observed in the area ranging from automation technology to transport infrastructure~\cite{zurawski2014industrial}. In comparison to traditional office Ethernet based networks, the modern Ethernet technology usage is much more challenging. Recent trends of technology exhibit the requirement of deterministic behavior of network. Amidst such demand, Profinet~\cite{feld2004profinet} is found to be an established real-time Ethernet based standard that is optimized for communication in automated industrial environments.

\begin{figure}[h]
\centerline{\includegraphics[width=8.5cm,height=4.5cm]{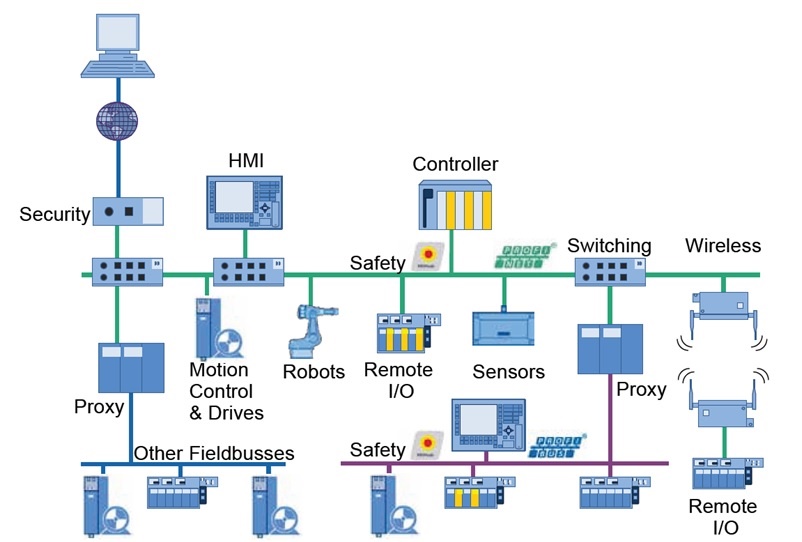}}
\caption{Profinet network in an automation system \cite{profinetArch}.}
\label{figure:profinetArch}
\end{figure}

Industrial automation systems are complex in nature and usually structured into  several hierarchical levels. Each of the level demands appropriate communication. These hierarchical levels can be categorized as 1) field-level networks, 2) control-level networks and 3) information-level networks~\cite{jasperneite2005profinet}.
At the  automation field  level, various sensors,  actuators,  I/O  modules and drive units  communicate via Profinet. Office- or information-level deals with centralized PCs for monitoring purposes. Automation control level includes controllers like PLC \cite{kleines2008performance} that communicate with actuators and other IO devices with industrial field buses Profibus as illustrated in Figure \ref{figure:profinetArch}.

\begin{figure}[t]
	\centering
	\vspace{-1mm}
    \includegraphics[width=0.8\linewidth, height=0.5\linewidth]{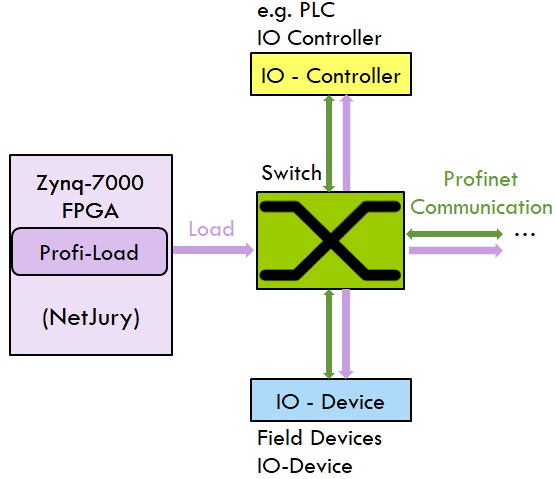}
	\caption{\textit{Profi-Load} in Nutshell}
	\label{figure:framework}
	\vspace{-2mm}
\end{figure}

For instance, position control of electric AC motors is a widely adopted application in industrial automation environment. In this specific case, the control system requires real-time response within a strict duration. Hence, analysis of some significant network performance is required to evaluate the accomplishments of this requirement. To evaluate the effectiveness of such applications, we need a rigorous testing environment, where we can observe the performance of the application under various network conditions. Real-time
characteristics and determinism issues, are the two uttermost important factors in Profinet based communication. To evaluate these two factors, one requires a framework which can generate specific network packets with stringent timing specification (commonly knows as \textit{net load} or \textit{Load generation}) \cite{netload}.

Recently, with the advent of new hardware based solutions, FPGAs have emerged as a bright prospect for computation sensitive applications~\cite{monmasson2017fpgas}. Due to its inherent parallelism application, execution speed on this platform is much faster than software-based execution on commercial-of-the-shelf (COTS) hardware. Reconfigurability is another interesting feature that FPGAs exhibit. Due to this unique feature, the same system can be reconfigured multiple times by loading new configuration information (bitstream) given different requirements.

In this paper, we have proposed an FPGA-based solution for load generation in the Profinet network. We coined our current traffic generation framework as ``\textit{Profi-Load}'' and is shown in Figure \ref{figure:framework}. In order to make the framework efficient, we have chosen the FPGA-based Xilinx NetJury device~\cite{netJuryXilinx}, as our implementation platform. The experimental validation of our proposed framework on the aforementioned device showed that \textit{Profi-Load} can effectively achieve required amount of load generation capacity with strict timing restrictions. The main contributions of this article can be stated as:
\begin{itemize}
    \item Defined a framework, ``\textit{Profi-Load}'' for generating specific amount of network load in real-time Profinet communication.
    \item Implemented the framework on an FPGA-based NetJury platform.
    \item The HMI based user interface on NetJury provides a friendly environment to the user such that specific timing and load generating parameters can be provided.
    \item \textit{Profi-Load} systematically exploits the reconfigurable feature of FPGA and thus, is capable to configure the NetJury device to assimilate user defined requirements. 
    \item The experimental validation and more specifically, Matlab, Hilscher netANALYZER and tektronix DPO4054B oscilloscope based analysis reveals the efficacy of our proposed framework on the NetJury device.
\end{itemize}

The remainder of the paper is organized as follows. Section II,
gives a brief background about the load generation strategy for industrial networks and related work. The \textit{Profi-Load} framework is discussed in Section III
by illustrating different user-defined cases. Section IV, provides a brief discussion of the NetJury device and its associated functionality. Section V, discusses on experimental outcome of the proposed framework with a  discussion on the same. Finally, the paper concludes with
Section VI.

\vspace{-3mm}

\section{Background \& Related Work}

\subsection{Industrial Network Load Generator}
Packet generator is a system which allows sending custom packets into the network with desired parameters specified by the user. Similarly, a network load generator can also be considered as a special kind of packet generator which considers a specific rate to send the packet(s) into the network in order to achieve the user-defined \% of load. In general, packet generators can be broadly categorized into two types, based on the platforms on which it can be implemented. One type of generator is software
based. There are many open source software-based packet generators including Pktgen~\cite{pktGen} , scapy~\cite{scapy}, ostinato~\cite{ostinato} and all of these can run either on Linux or Windows given console or GUI.  Another type is the \textit{hardware-based packet generator}, this hardware-based solution involves processing elements like FPGAs, ASIC etc.

The basic hurdles for  software-based solutions are its restriction on full line-rate testing. For example,  proprietary products such as Ixia~\cite{ixia} are expensive and are not directly available to the open-source community. As an alternative, FPGA-based NetJury device allows an open-source implementation of packet generator coupled with the packet receiver operated at Gigabit/Fast Ethernet line rates. However, such hardware or software based solutions while generating packets, do not consider the amount of load to be maintained in the network, therefore, our proposed \textit{Profi-Load} targets to generate traffic for the desired load as specified by the user. 

\subsection{Related Work}

In recent past, the research community has focused on developing and improving various aspects of industrial automation systems and Profinet played a pivotal role here \cite{antolovic2006plc}. Researches have spun off in various directions to incorporate various advance modification on Profinet performance. In~\cite{dias2017performance}, authors depicted the performance analysis of
Profinet communication applied in a 
motion control application. In~\cite{ferrari2011large}, an actual case study regarding a real
Profinet IO network for factory automation has been discussed. The outcomes from their experiments showed that Profinet is able to maintain strict timeliness even when the device is heavily loaded. 

A communication scheduling and remote estimation problem studied in \cite{gao2019communication}. Given the sensory element, the authors have employed an encoder-decoder strategy coupled with a feedback Stackelberg solution to minimize the overall communication cost. To deal with the energy depletion and hardware malfunctioning in wireless sensor-based industrial network, authors in \cite{duan2017methodology} proposed a software based framework, named as SD-WSN. The software-based SD-WSN architecture successfully addresses the problem of node failure, network reliability and management. Employment of FPGAs in the domain of industrial automation can be observed in\cite{durkop2012towards}. Here, the embedded processor (Nios) inside the FPGA has been used to assimilate a Profinet software IP core. However, this work still belongs to the software-based implementation category as it did not involve FPGA fabrics for implementation.  

First attempt to devise a hardware based Profinet implementation can be found in~\cite{flatt2012fpga}. This FPGA-based implementation reveals the fact that the proposed architecture is re-configurable and thus, suitable for  high-end industrial requirements. Recently, in~\cite{nguyen2018fpga}, the authors have shown the FPGA-based implementation of  Media  Access  Control  (MAC)  and  Physical  layer  system (PHY)  for  industrial automation   systems. From the above discussion, it can be established that FPGAs are evolving as a promising platform for industrial automation devices. However, an FPGA-based solution which provides a user friendly access to generate specific network conditions is still in its infancy. Thus, in this paper, we have explored the FPGA-based NetJury device to devise a load generator for Profinet communication. The next section will elaborate our proposed \textit{Profi-Load} framework.

\section{Profi-Load Framework}
\label{sec:profiLoad}
In this section, we will describe how our proposed \textit{Profi-Load} framework will function. In a nut-shell, the proposed framework generates packets (frames)\footnote{In this paper, we have used the terms \textit{packets} and \textit{frames} interchangeably} and place inter-frame gaps in such a way that the desired load can be maintained. Before describing the core theme of our strategy, we will now discuss the default parameters for generating the packet.

\subsection{Parameters of a Packet}
\label{sec:packet}
\begin{figure}[htbp]
\centerline{\includegraphics[width=8.5cm,height=3cm]{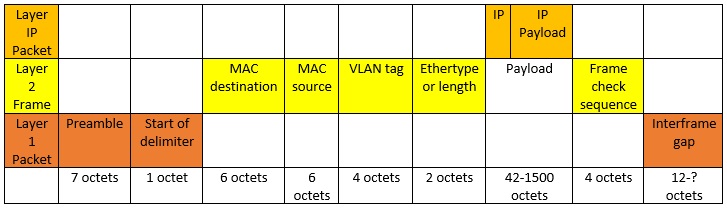}}
\caption{Structure of a packet.}
\label{figure:packet}
\end{figure}

A network packet contains header fields which includes source and destination MAC addresses and is denoted as $M$. A packet also contains Ethertype (denoted as $E$) i.e., for Profinet packet, Ethertype would be ``0x8892'', optional vLAN tag (denoted as $v$; vlan id, vlan priority and vlan cfi) and payload size (denoted as $Y$) in bytes. MAC addresses $M$, Ethertype $E$ and Payload $Y$ are collectively represented as $P$ in equation~(\ref{eq:0}). For ease of visualization, a basic packet structure is shown in Figure \ref{figure:packet}. Apart from these main parameters, a packet also contains some other parameters including preamble: $p$, delimiter: $d$, frame check sequence: $f$  and interframe gap: $i$. All these parameters are combined with the packet header fields to represent a single frame. We denote all such other fields as $O$ overhead in equation~(\ref{eq:1}). If the packet is $v$ vLAN tagged then an extra $4$-bytes get included in the overhead $O$.

\vspace{-1mm}
\begin{equation}
    P = \{M, E, Y\}
    \label{eq:0}
\end{equation}
\vspace{-1mm}
\begin{equation}
    O = \{p, d, f, i, v\}
    \label{eq:1}
\end{equation}

The overall size of a packet, $S$ will include both $O$ and $P$ in bytes as shown in equation~(\ref{eq:2}). Varying payload $Y$, $P$ is of size 60, 128, 256, 512, 1020, 1514 bytes.

\begin{equation}
    S = \{P, O\} 
    \label{eq:2}
\end{equation}

\subsection{Load Generation Strategies}
\label{sec:profiLoadStratrgies}
In this subsection, we will discuss the strategies adopted by the \textit{Profi-Load} framework. As per the requirements of modern industrial automation devices, the load generator should be able to generate load with some specific constraints i.e., either by selecting the number of frames or the time duration. \\

\noindent
\textbf{CASE 1: Sending $F$ Number of Frames for a Specified Load}\\
Let us assume the case where \textit{Profi-Load} has to send $F$ number of frames with $L$ \% of load. Here, as the number of frames is fixed so in order to maintain the fixed ($L\%$) load, \textit{Profi-Load} will send frames with interframe gap ($I$). Hence, a certain gap between two consecutive frames should be maintained. This $I$ can be calculated as:
\begin{equation}
    I_L = S \times (\frac{1}{L}-1)
    \label{framefix1}
\end{equation}

\vspace{-1mm}
\begin{equation}
    I = I_d + I_L
    \label{framefix2}
\end{equation}

\begin{figure}[htbp]
\includegraphics[width=1.0\linewidth, height=0.1\linewidth]{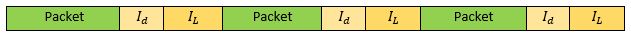}
\caption{Interframe gap between three consecutive frames to maintain desired load in the network.} 
\label{figure:interFrameGap}
\end{figure}

In the equation~(\ref{framefix2}), $I_d$ denotes the minimum interframe gap of $12$ bytes between two frames. However, to maintain the desired percentage of load, $I_L$ is the additional interframe gap in bytes to be added while sending specific number of consecutive frames (say, $3$) as shown in Figure \ref{figure:interFrameGap}. This scenario will be effective when the user is interested to employ \textit{Profi-Load} for load generation by specifying a particular number of packets.

\noindent
\textbf{CASE 2: Generating Load for $T$ Time duration}\\
Let us assume another scenario within which the user needs the \textit{Profi-Load} to generate  $L$ \% of load for the specified duration of $T$ time units\footnote{time units could be hours, minutes or even micro seconds as per the requirements}. Therefore, in $T$ time period, \textit{Profi-Load} will calculate and send $F$ number of frames which will be passed through the Profinet network so that the required $L$ \% load can be achieved. The required $F$ number of frames can be calculated as:

\begin{equation}
     F = \frac{R}{S \times 8} \times L \times T
     \label{timefix}
\end{equation}

In equation (\ref{timefix}), $R$ denotes the available network bandwidth which can either be 100 Mbps (Fast Ethernet) or 1 Gbps (Gigabit Ethernet).
However, as per our experimental observation, it has been observed that the elapsed time (T') i.e., the actual time consumed for sending  $F$ number of frames (obtained from equation~\ref{timefix}) can be equal or slightly lower (of the order of microsec) than T. We will analyze this phenomenon in detail (refer, Section~V, \textit{Evaluation of Case 2}).\\

\noindent
\textbf{CASE 3: Generating Bursts of Specified Load}\\
Given the user defined $F$ number of frames and $T$ time period, case 1 and 2 will result into a non-synchronized load generation. In real-time industrial environment, such load generation might not be efficient due to variable response timings of the IO devices set by the PLC controllers \cite{feld2004profinet}. It might eventually cause the industrial devices to enter into the failure mode which is not desirable. To keep the devices under normal operation, \textit{Profi-Load} should be able to generate load (packets) with specific delays. Therefore, the third scenario specifically works in this direction, group of packets (burst) are collectively send at some desired load for some specific time interval. After sending the required number of packets (burst), there will be some time-gap after which, the next burst will be transferred. The time duration in which the burst of packets are being transmitted is called ``Burst interval'' and the time-gap in between two bursts of packets is considered as a ``Sleep Interval''.





\section{\textit{Profi-Load} on FPGA (NetJury Device)}

\subsection{Brief Description of Netjury}
\begin{figure}[htbp]
\centerline{\includegraphics[width=8cm,height=5cm]{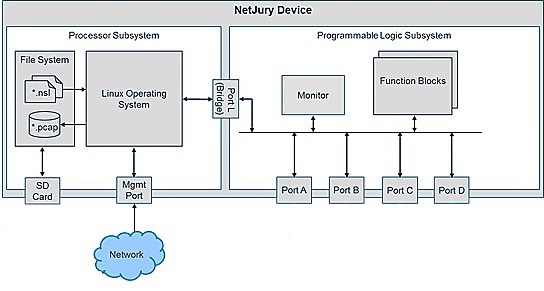}}
\caption{Architectural view of the NetJury~\cite{netJuryXilinx}.}
\label{figure:NetJury}
\end{figure}

As the execution platform of our proposed \textit{Profi-Load} framework, we have selected the NetJury device~\cite{netJuryXilinx}. NetJury is based on Xilinx Zynq FPGA~\cite{rajagopalan2011xilinx} integrated with a Dual-core ARM Cortex-A9 processor~\cite{wang2011survey}. The FPGA fabric part is termed as PL (Programmable Logic) and an ARM component named as the Processing System (PS). PL is mainly responsible for hardware processing of real-time data and the PS is used to operate the PL logic unit via Linux based environment. NetJury supports a proprietary scripting language, named as NetJury Scripting Language (NSL) to program the PL component. The overall architecture of the device is shown in Figure \ref{figure:NetJury}.    

In Figure \ref{figure:NetJury}, the right block is the programming part (PL) and the left block (PS) provides the Linux environment on the ARM processor. The PL subsystem consists of four Ethernet ports operated at Gigabit Ethernet (1Gbps) and Fast Ethernet (100Mbps). Within the scripting mode, packet generation with manipulation up-to the application layer can be carried out with all the Ethernet ports via NSL script, operated from the PS block. The Port L (eth1) is used to communicate between PS and PL. Monitor module helps the user to inspect and manipulate the traffic received on any of the FPGA ports. The user from the PS side can trigger the desired functionality in PL with NSL scripts generated and executed by high level programming language e.g. python. However, the user needs to turn on the scripting mode once at start for using the PL side of the NetJury. 


\subsection{\textit{Profi-Load} inside NetJury: Human Machine Interface}

NetJury by default is designed to send traffic while expecting a particular number of frames to be received on any of the desired port. Hence, to employ the \textit{Profi-Load} framework on NetJury, the user needs to provide all the required information. 
In order to generate the load  with certain specifications (as discussed in previous section via different cases) the user has to provide the following information (based on the requirements):
\begin{itemize}
    \item \textit{Load} \%
    \item \textit{Source MAC address}
    \item \textit{Destination MAC address}
    \item \textit{Ethertype}
    \item \textit{vLAN fields}
    \item \textit{Packet size}
    \item \textit{Load generating port of NetJury and its line rate}
    \item \textit{Functionality}
    \begin{itemize}
    \item \textit{Burst feature}
    \item \textit{Time feature}
    \item \textit{Frame feature}    
    \end{itemize}

\end{itemize}

\begin{figure}[htbp]
\includegraphics[width=0.9\linewidth, height=1\linewidth]{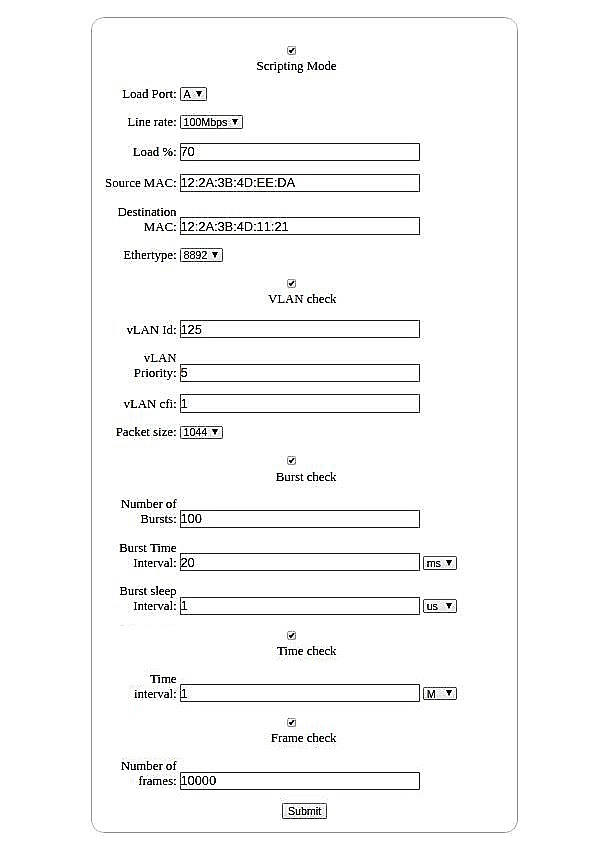}
\caption{ Load generation interface for user input.}
\label{figure:net_load}
\end{figure}

Upon receiving the user input, NetJury will calculate $F$ number of frames and $I$ inter-frame gap (for each consecutive frames) for load generation. 
To access our \textit{Profi-Load} framework efficiently, we have developed a web-based Human Machine Interface (HMI). This GUI interface will provide NetJury the required information for successful load generation shown in Figure \ref{figure:net_load}. Once the user successfully logged in, 
the GUI will allow the user to provide the input arguments. However, the user can only make use of single feature at a time, i.e. it can either be Burst check, Time check or Frame check. Once the load is generated successfully, via HMI, NetJury will inform the user about the actual amount of feasible frames or time spent to generate required \% of load.

\begin{figure}[htbp]
\centerline{\includegraphics[width=8cm,height=6cm]{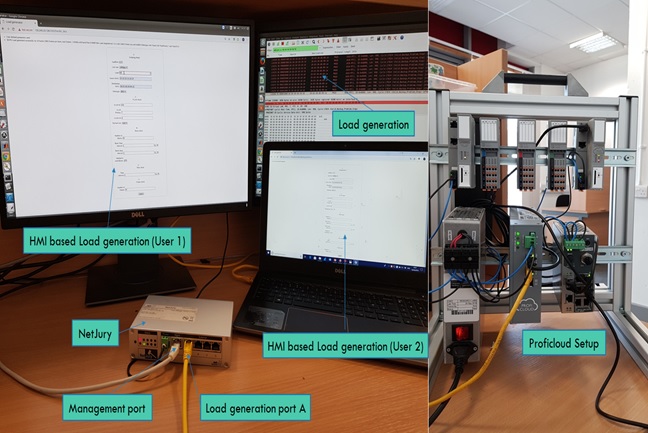}}
\caption{\textit{Profi-Load} setup.}
\label{figure:profiLoadSetup}
\end{figure}

\section{Experimental Validation}
In this section, we detail the measurements that were carried out to test the efficacy of our \textit{Profi-Load} framework for real-time traffic generation. Inside the NetJury, the synthetic packets are created by using pythonic modules such as Scapy and translated into NSL syntax for load generation. These packages are executed in Linux environment running on the PS side. As measurement tools, we have used Wireshark~\cite{orebaugh2006wireshark},  Hilscher netANALYZER~\cite{HilscherAnalyzer} and a tektronix DPO4054B Oscilloscope~\cite{tektro}. Moreover, accurate time measurements with Ethernet decoding are performed via MATLAB. 

Figure \ref{figure:profiLoadSetup} illustrates the running setup of \textit{Profi-Load} employing NetJury. The management port of the NetJury is connected with the Intra-net from where any connected device such as Laptop and other mobile devices (say, user 1 and 2) with user credentials can access the HMI interface for load generation. Wireshark measurement is carried out by connecting the packet generating port (port A as shown in Figure \ref{figure:profiLoadSetup}) of the NetJury to the Ethernet port of a PC running Wireshark application. Depending upon the OS level interrupt(s), the captured timestamps through Wireshark can vary from real-time measurements performed with Hilscher netANALYZER and tektronix DPO4054B oscilloscope. Now we will describe our experimental observations on the performances of \textit{Profi-Load} over different cases.

\subsubsection{ Evaluation of Case 1: (40 \textbf{long} frames at 25\% load)}
The frames are being sent with $P = 1514$ bytes. Alongside, there will be an extra 7 bytes preamble, 1 byte start delimiter, 4 byte frame checksum and for Fast Ethernet, a default 12 bytes interframe gap. Therefore, according to equation~(\ref{eq:1}), the overhead ($O$) is 24 bytes and the total size of the packet becomes $S = 1538$ bytes (refer equation~\ref{eq:2}).

To send 40 packets with 25\% of load, according to equation~(\ref{framefix1}), $I_L$ becomes 4614 bytes and $I_d$ is 12 bytes. Hence, interframe gap ($I$) that we should maintain  becomes 4626 bytes (refer, equation~\ref{framefix2}).

\begin{figure}[h]
\centerline{\includegraphics[width=8.5cm,height=5cm]{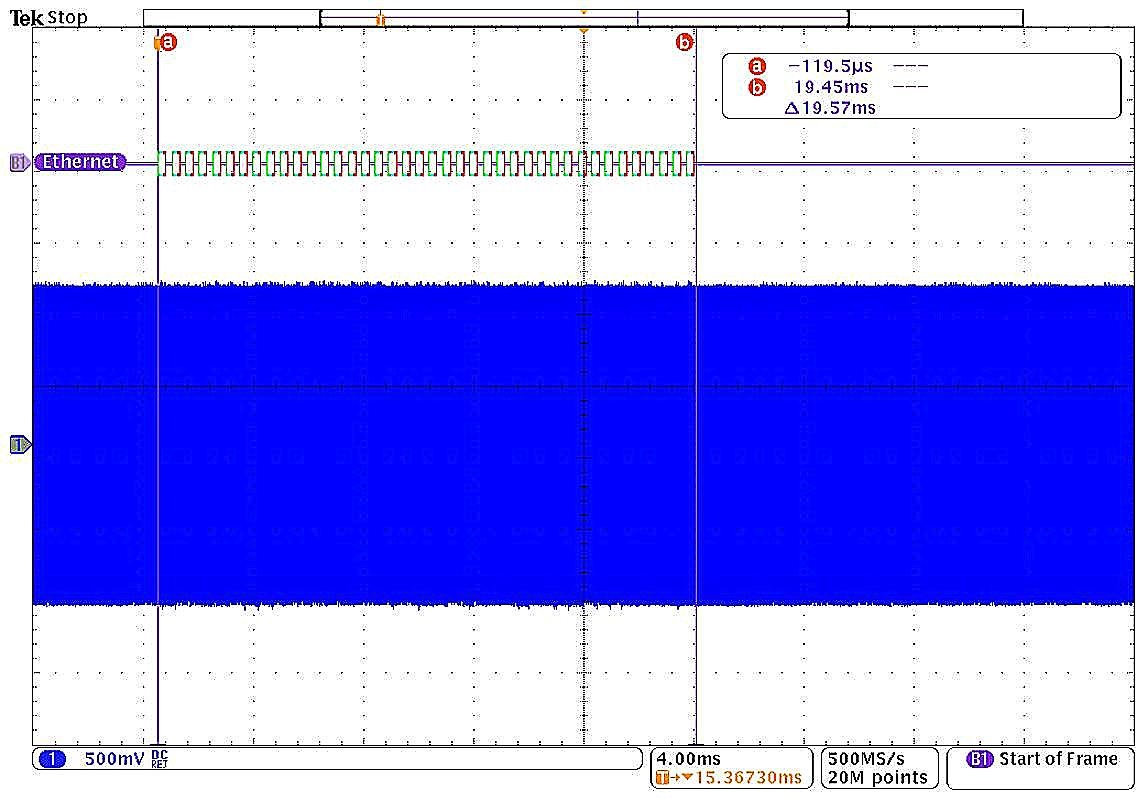}}
\caption{Oscilloscope measurement of 40 long packets at 25\% load.}
\label{figure:40long_osil}
\end{figure}

\begin{figure}[h]
\centerline{\includegraphics[width=8.5cm,height=6cm]{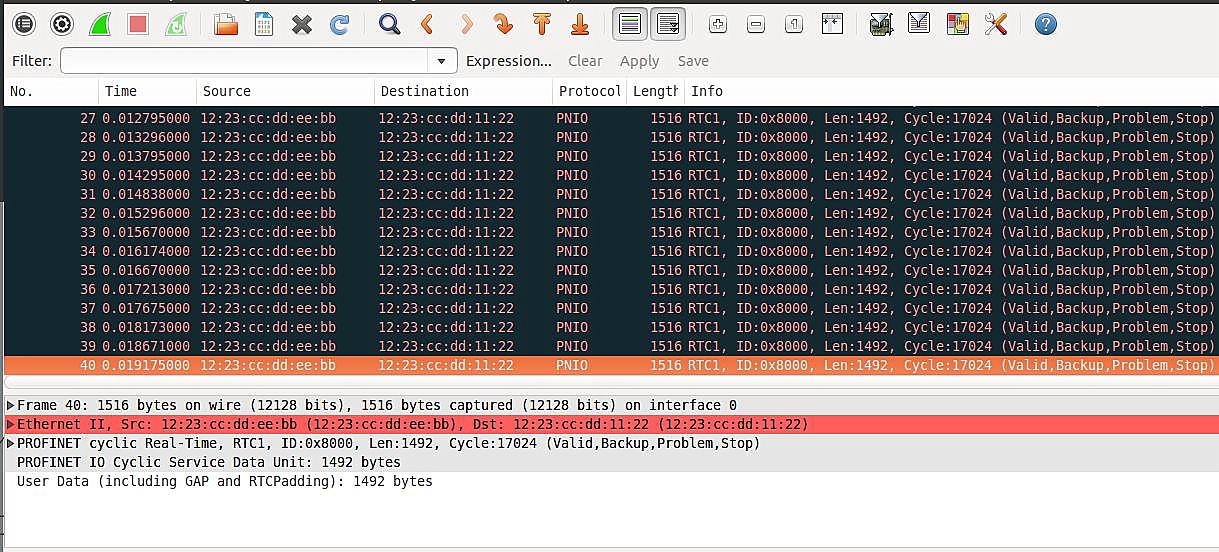}}
\caption{Wireshark measurement of 25\% load with 40 long packets.}
\label{figure:25_40LongFrames}
\end{figure}

After receiving all the input parameters via HMI interface, NetJury will calculate the $F$ number of frames, $I$ interframe gap and other parameters and will start operating. Figure~\ref{figure:40long_osil} shows the Oscilloscope measurement of 40 long packets. It has been observed that the entire transmission of the 40 packets took 19.57 ms. We have also employed Wireshark to measure the transmission time and observed 19.17 ms for the transmission (please see Figure~\ref{figure:25_40LongFrames}). \\ 

\begin{figure}[h]
\centerline{\includegraphics[width=8.5cm,height=4cm]{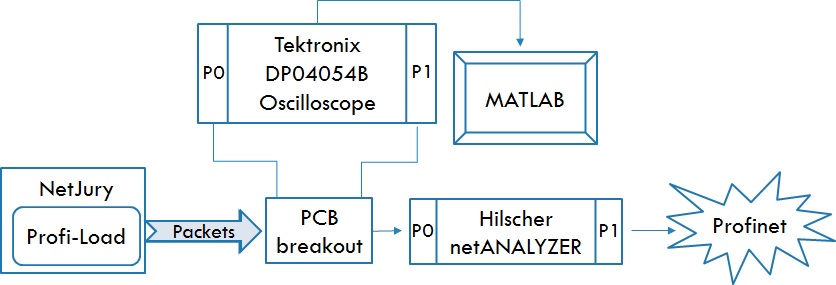}}
\caption{Setup for capturing packets using tektronix DPO4054B Oscilloscope.}
\label{figure:netJury_oscil_analyzer}
\end{figure}

\noindent
\textbf{Matlab Based Analysis:}
 Using Oscilloscope, we captured the \textit{Profi-Load} generated packets with setup shown in Figure \ref{figure:netJury_oscil_analyzer}. P0 and P1 depict the probes / ports of the Oscilloscope / Hilscher netANALYZER coupled with a PCB board specifically made for connecting probes and ports. Each captured packet has been further evaluated via MATLAB. For this evaluation, we have introduced two new parameters: $E_R$ which denotes packet lasting time in the Profinet network and $E_L$ which denotes the specific time-gap after which each frame will be sent periodically to maintain the specified load $L$.

$E_R$ can be found out for $R=100 $ Mbps link, as:
\begin{equation}
    E_R = \frac{S \times 8} {R}
   \label{er}
\end{equation}

After finding the $E_R$, we can find the $E_L$ as:
\begin{equation}
    E_L = \frac{E_R}{L}
     \label{el}
 \end{equation}
 
 Theoretically, with the above two equations (equation~\ref{er} \& equation~\ref{el}), we can expect the values of $E_R= 123.04 \mu s$ and $E_L= 492.16 \mu s$ for $L=25\%$ load.
 
\begin{figure}[h]
\centerline{\includegraphics[width=8.5cm,height=6cm]{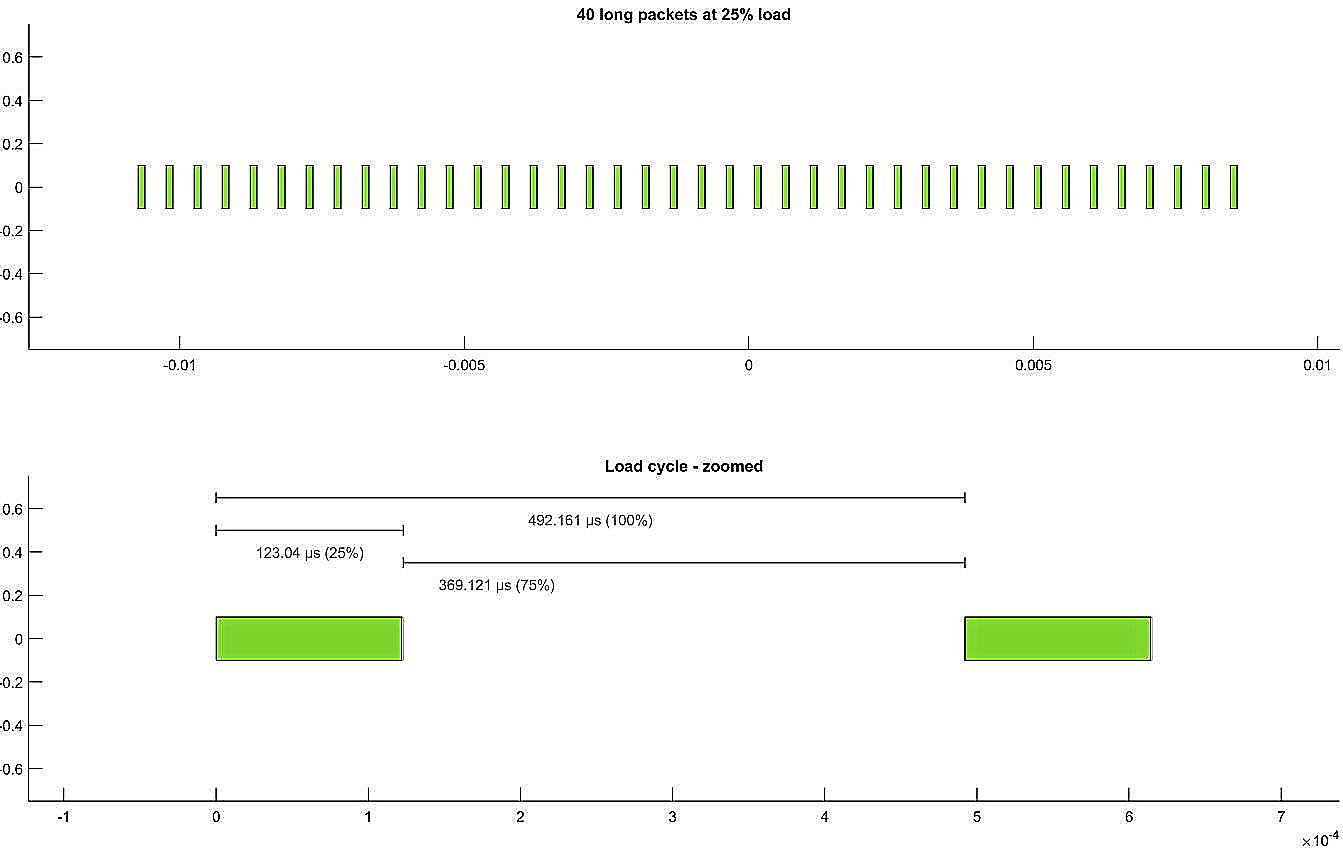}}
\caption{MATLAB analysis of 40 long packets at 25\% load.}
\label{figure:40long_osil_mat}
\end{figure}

 Experimentally, we have achieved the same results. In Figure~\ref{figure:40long_osil_mat}, it can be observed that NetJury sent 40 packets with 25\% load. A single packet lasts as expected for 123.04 $\mu s$ and the NetJury sent each packet (upto 40 packets) after a specific time gap ($E_L$) of 492.161 $\mu s$. From the above experimental evaluation, it is evident that NetJury can effectively generate load with proper timing efficiency.



\begin{figure}[h]
\centerline{\includegraphics[width=8.5cm,height=6cm]{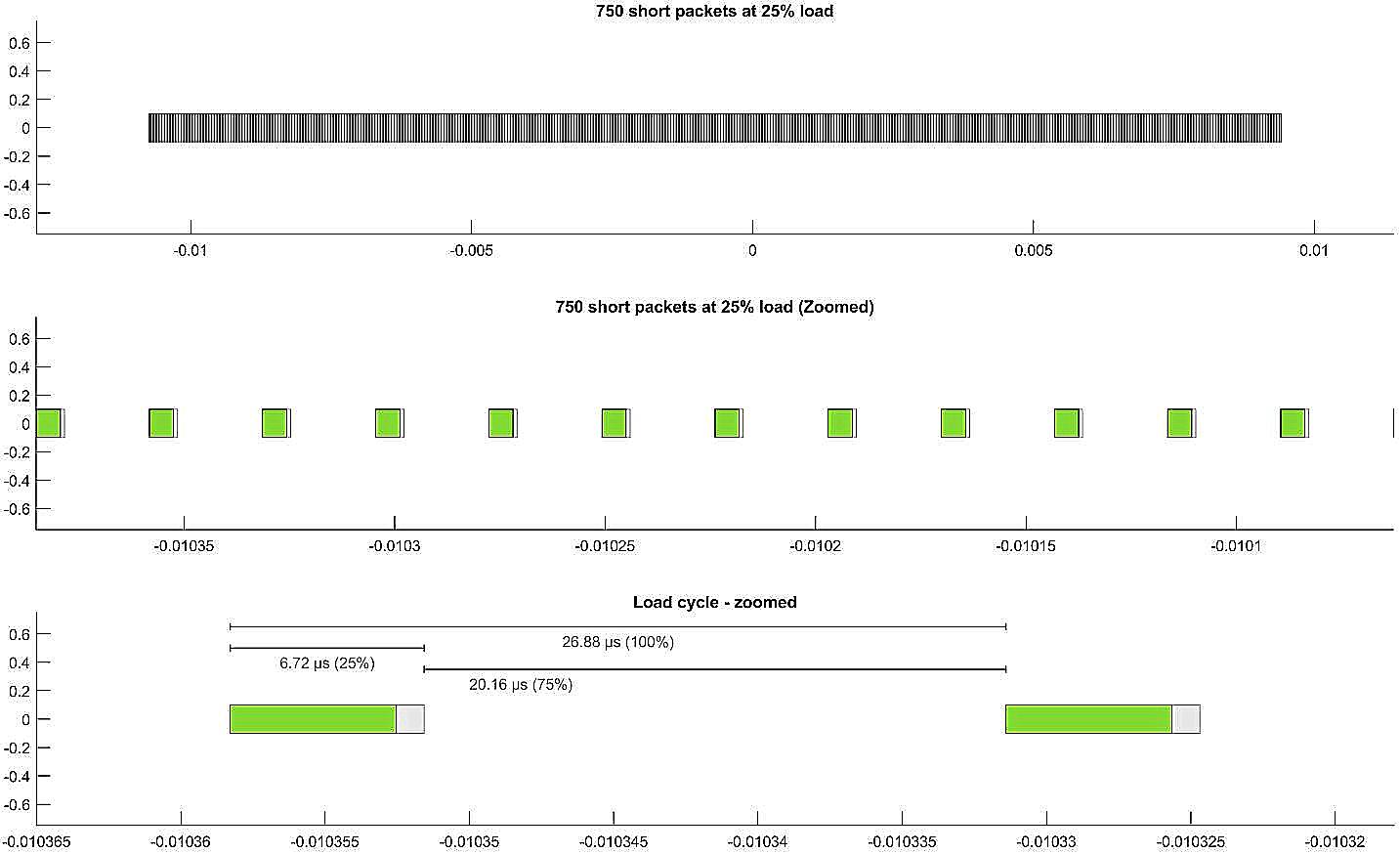}}
\caption{MATLAB analysis of 750 short packets at 25\% load.}
\label{figure:750short_osil_mat}
\end{figure}

\begin{figure*}[t]
\centerline{\includegraphics[width=11.4cm,height=6.4cm]{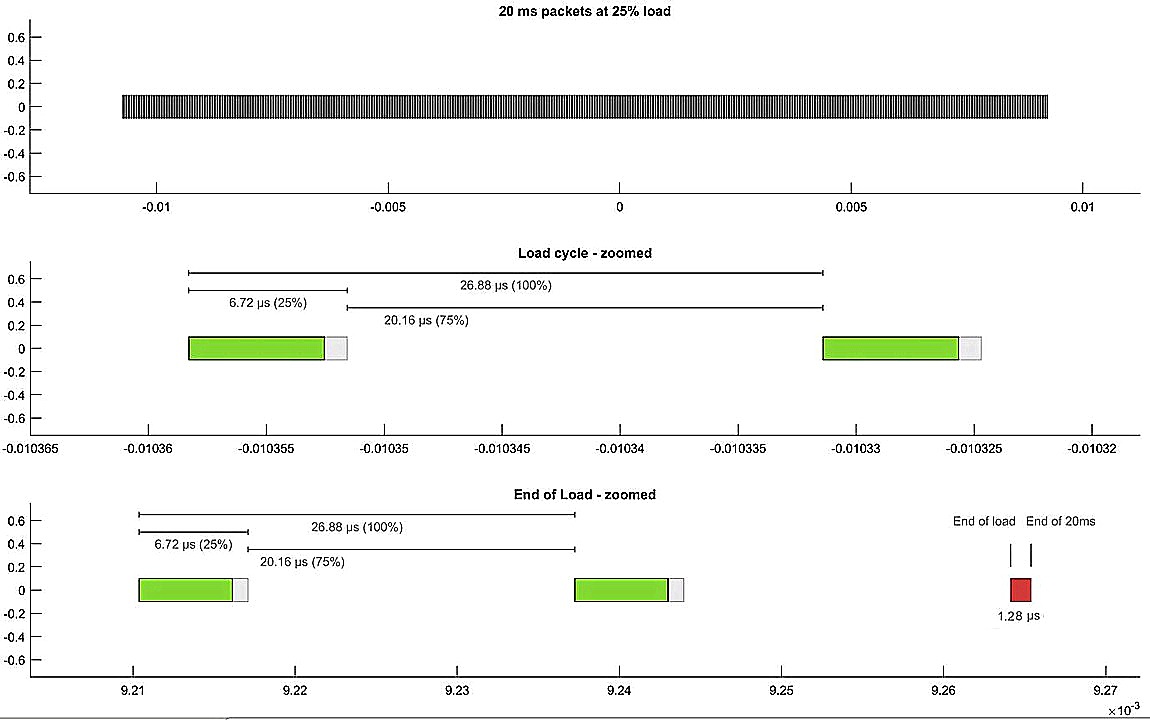}}
\caption{MATLAB analysis of 25\% load with 744 short packets in 20 ms.}
\label{figure:20msshort_osil_mat}
\end{figure*}

 \begin{figure}[h]
 \centerline{\includegraphics[width=8.5cm,height=6cm]{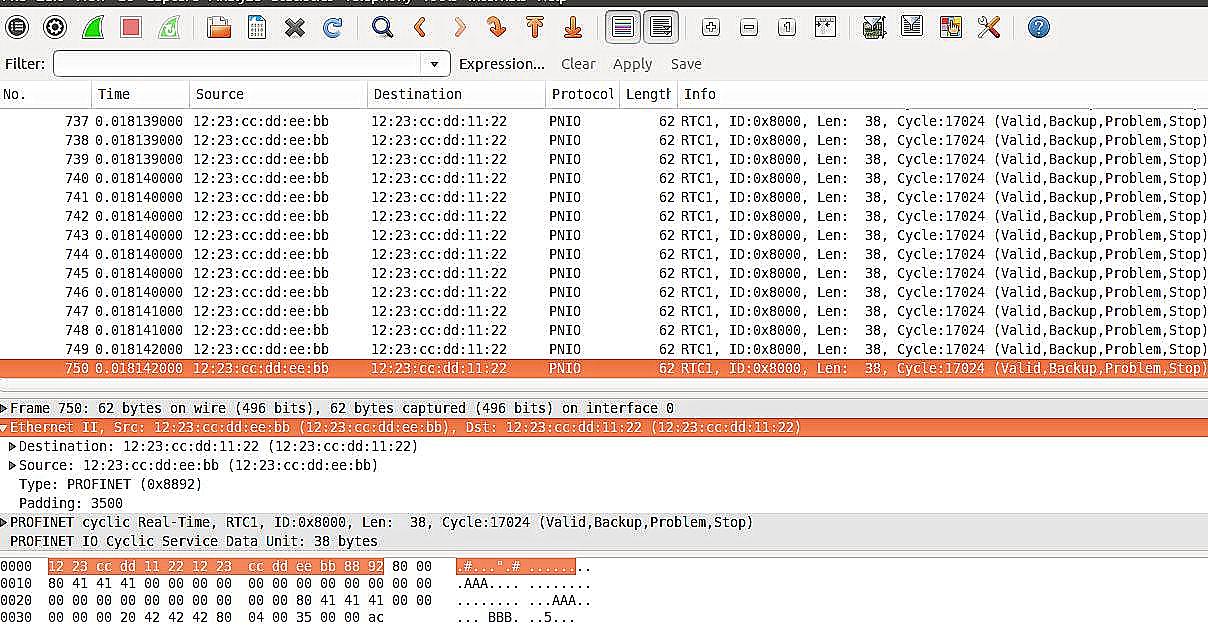}}
 \caption{Wireshark measurement of 25\% load with 750 short packets.}
 \label{figure:25Per_750ShortFrames}
 \end{figure}

\subsubsection{ Evaluation of Case 1: ($750$ \textbf{short} packets at 25\% load)}

It has also been observed that \textit{Profi-Load} is capable of performing efficiently in case of generating short frames with a specified load. In this case, the short frame size ($S$) is 84 bytes, where $P$ is of $60$ bytes and $O$ contains $24$ bytes.

In Figure~\ref{figure:25Per_750ShortFrames}, it can be seen that Wireshark detected that \textit{Profi-Load} generates 750 short frames  in $18.14$ ms. A similar Matlab based analysis revealed the fact that NetJury generates 25\% load by sending every frame with the interval of $E_L = 26.88\mu s $. Each packet lasts for $E_R = 6.72\mu s$ (refer, Figure~\ref{figure:750short_osil_mat}). These measurements are also matched with the theoretically expected values of $E_R$ and $E_L$.




\subsubsection{Evaluation of CASE 2: (Short packets at 25\% load for $20$ ms \textbf{time period)}}
After receiving the specific inputs from the user, the \textit{Profi-Load} framework will calculate the required number of frames that need to be sent for maintaining the load ($L\%$) for specific duration ($T$).

In this particular case, the size ($S$) of each frame is 84 bytes ($P=60$ bytes and $O=24$ bytes respectively). The speed of the network ($R$) is 100 Mbps. The required load ($L$) is 25\% and  the specified time duration ($T$) is 20ms. Hence, as per equation~(\ref{timefix}), the number of required frames ($F$) which have to be sent, comes around to be $744$.

These $744$ frames/packets have to be sent with the specific interframe gap ($I$) which can be calculated from equation~(\ref{framefix2}). Here, $I$ can be found as
\begin{equation*}
   I = 12+252 = 264 
\end{equation*}

As per our previous measurement strategies, Fig.~\ref{figure:20msshort_osil_mat} illustrates the Matlab based analysis. It has been observed that the NetJury can generate frames with correct / expected specifications. Each frame lasts for $E_R= 6.72 \mu s$ and the $E_L = 26.88\mu s$ time gap, after which the next frame must be sent. Thus, by following this interval period ($E_L$), $744$ packets / frames were sent for $T = 20 ms$.\\

\noindent
\textbf{Critical Observation and Analysis:}

We have previously mentioned in section \ref{sec:profiLoadStratrgies} (Case~2) that the user specified duration ($T$) can slightly vary from the experimentally observed duration ($T'$). Here, we can calculate the time elapsed ($T'$) for sending $F$ number of frames as:

\begin{equation}
   T' =  {E_L \times F} 
   \label{at}
\end{equation}

From (\ref{at}), the actual elapsed time ($T'$) for sending 744 frames becomes $19.99872 ms$. From Figure~\ref{figure:20msshort_osil_mat}, it can be observed that this time difference (say, $TD$) as:
\begin{equation*}
    TD = 20-19.99872=1.28 \mu s
\end{equation*}

This observation can be attributed to the fact that for the specific duration of 20 ms, \textit{Profi-Load} found it feasible to send only 744 packets to maintain 25\% load despite having a small $TD$. However, in order to compensate $TD$, if \textit{Profi-Load} sends an extra frame then the elapsed time ($T'$) would exceed the specified duration ($T$) as:
\begin{equation*}
    T' = 19.99872 + E_R > 20 ms~ (T)
\end{equation*}
Hence, this observation shows how effectively the NetJury device operates for generating specified load with stringent time requirements.




\begin{figure}[h]
\centerline{\includegraphics[width=8.5cm,height=6cm]{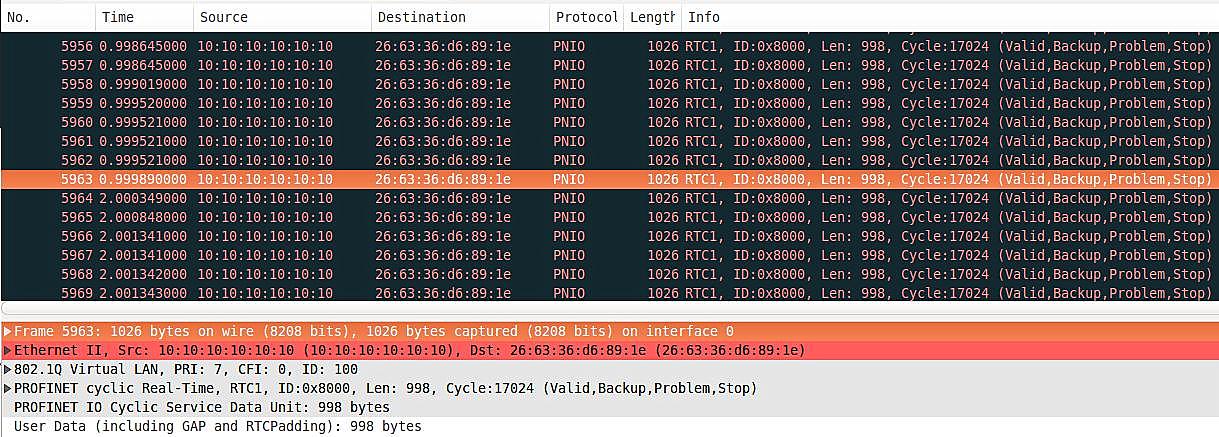}}
\caption{Wireshark measurement of 25\% load exhibiting the end of $1^{st}$ Burst and start of $2^{nd}$ Burst with $1s$ sleep time.}
\label{figure:50Perc_1stBurst}
\end{figure}

\begin{figure}[h]
\centerline{\includegraphics[width=8.5cm,height=6cm]{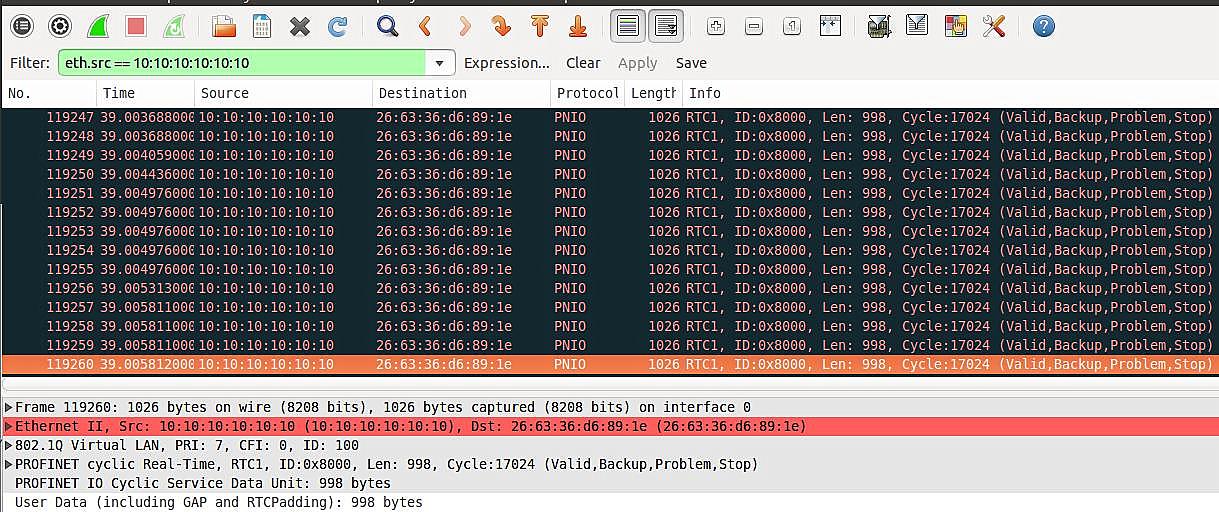}}
\caption{Wireshark measurement of 25\% load exhibiting the total number of frames generated by 20 bursts.}
\label{figure:allPkts20Bursts}
\end{figure}

\subsubsection{Evaluation of CASE 3: (\textbf{vLAN} tagged 50\% load for $20$ \textbf{bursts})}

In modern factory automation environment, different I/O devices demand different synchronizing/response times. Therefore, injecting a non-synchronous load (via case 1 and case 2) may force them to enter into the failure mode, thus, devices will stop functioning. Therefore, it will be effective to generate load with ``burst feature". This feature will enable user to specify number of bursts (each burst will contain a specific number of packets) to maintain the desired load. The time duration between two consecutive bursts is referred here as ``Sleep Interval''.



Consider a scenario where the user would like to send 50\% vLAN tagged Profinet load with 20 bursts. Both the ``Burst Interval'' and ``Sleep Interval'' are kept as $1$ second. With $P=1020$ bytes and having
an extra 4-bytes $v$ vLAN tag, the overhead becomes $28$ bytes. Hence, the packet size ($S$) is of $1048$ bytes.

After obtaining all these user specified parameters, \textit{Profi-Load} will calculate the $F$ number of frames to be sent using equation~(\ref{timefix}). It has been found that for maintaining 50\% load there should be $5963$ number of frames within a burst.
Thus, $20$ bursts will collectively generate around $119260$ frames.

If we look into the Figure~\ref{figure:50Perc_1stBurst}, we can identify that the $1^{st}$ burst ends at $5963^{th}$ frame. After the specified sleep interval of $1 s$, 
$2^{nd}$ burst start sending frames. It can also be observed that the \textit{Profi-Load} can transmit \textit{VLAN} tagged packets as well with specified priority (7, in this case). Figure~\ref{figure:allPkts20Bursts} establishes the fact that the \textit{Profi-Load} successfully generates the required number of bursts with specified duration. It is evident that total $119260$ frames were sent within 39 seconds i.e 20 seconds for transmission and 19 seconds for the sleep interval. 

Currently, the burst feature keeps the burst interval or sleep interval same for all the user defined number of bursts. However, this feature can be extended for variable burst interval or sleep interval, i.e., $1s$ sleep interval between $1^{st}$ and $2^{nd}$ burst and $1ms$ sleep interval between $2^{nd}$ and $3^{rd}$ burst with burst intervals of $1.5 s$, $10  ms$ and $30 \mu s$ for $1^{st}$, $2^{nd}$ and $3^{rd}$ burst. Moreover, our proposed \textit{Profi-Load} can also be employed to generate load other than the Profinet-based packets by changing the Ethertype along with a suitable payload size (please see Figure \ref{figure:net_load}).



\section{Conclusion}

The vast deployment of Profinet in modern industrial automation devices such as, Controllers and the associated IO Devices can lead to unknown and untested critical failures when network traffic flows unexpectedly higher than normal. In this paper, we have proposed an FPGA-based  Profinet load generation framework, \textit{Profi-Load}. This framework can generate desired load configurations such as, bits per second, the number and size of the packets and the foremost factor delay in between the packets or burst of packets to maintain the desired throughput. Comprehensive set of experiments on a Zynq-7000 FPGA based ``NetJury" device, have showed that the proposed framework can effectively generate load under various user defined scenario. Moreover, the framework can easily be accessed via web based interface for successful load generation.   

\section*{ACKNOWLEDGMENT}

We gratefully acknowledge the support of Frederic Depuydt (Project Engineer / Researcher at KU Leuven, Technologiecampus Gent), the INCASE project partners for sharing knowledge in carrying out measurements with Hilscher netANALYZER and tektronix DPO4054B Oscilloscope used in this research.

\bibliographystyle{ieeetr}
\bibliography{root}
\end{document}